\documentclass[sigconf]{acmart}

\AtBeginDocument{%
  }

\setcopyright{acmlicensed}
\copyrightyear{2018}
\acmYear{2018}
\acmDOI{XXXXXXX.XXXXXXX}
\acmConference[Conference acronym 'XX]{Make sure to enter the correct
  conference title from your rights confirmation email}{June 03--05,
  2018}{Woodstock, NY}

\acmISBN{978-1-4503-XXXX-X/2018/06}

\usepackage[most]{tcolorbox} 

\usepackage{amssymb} 
\usepackage{marvosym} 
\usepackage{multirow} 
\usepackage{tabularx}
\usepackage{booktabs}
\usepackage{tabularx}
\usepackage{array}

\usepackage{rotating}
\usepackage{amsmath}

\usepackage{xcolor}

\renewcommand\footnotetextcopyrightpermission[1]{}
\definecolor{panelBack}{HTML}{F1F4F7}
\definecolor{panelRule}{RGB}{130,136,145}
\newtcolorbox{topmatterBox}{
    enhanced,
    colback=panelBack,
    colframe=panelBack,
    boxrule=0pt,
    frame hidden,
    arc=3mm,
    boxsep=0pt,
    left=6mm, right=6mm, top=5mm, bottom=4mm,
    before skip=4mm, after skip=0mm,
    notitle
}
\newcommand{\headerlogo}{%
  \includegraphics[trim={32.8 43.5 23.4 42.3}, clip,
                   height=4mm]{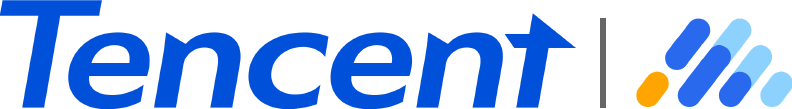}}

\newcommand{\panelnamefont}{\normalsize\bfseries}
\newcommand{\panelaffilfont}{\small}
\newcommand{\panelmailfont}{\ttfamily\footnotesize}
\makeatletter
\newbox\DASH@titlebx 
\let\@mktitle\@mktitle@i
\def\@titlefont{\Huge\bfseries}
\def\@mkauthors{\global\setbox\DASH@titlebx=\box\mktitle@bx} 
\def\@mkteasers{%
  \global\setbox\mktitle@bx=\vbox{%
    \hsize=\textwidth \linewidth=\textwidth \columnwidth=\textwidth
    \noindent\headerlogo\par
    \vskip 1.25mm 
    {\color{panelRule}\hrule height 1.2pt}%
    \vskip 4.75mm 
    \box\DASH@titlebx 
    \begin{topmatterBox}
      \setlength{\parindent}{0pt}
      \authorpanel               
      \medskip
      \noindent\ignorespaces\@abstract\par  
      \ifx\@keywords\@empty\else
        \medskip\noindent{\bfseries\keywordsname:} \@keywords\par
      \fi
      \medskip
      {\color{panelRule}\hrule height 0.4pt}%
      \smallskip
      \noindent{\small\authornotepanel}\par
    \end{topmatterBox}
    \vskip 3.5mm 
  }%
  \global\let\DASH@concepts\@concepts 
  \global\let\DASH@keywords\@keywords 
  \global\let\@mkabstract\@empty
  \global\@ACM@printccsfalse
  \global\let\@concepts\@empty
  \global\let\@keywords\@empty}
\let\DASH@printendtopmatter\@printendtopmatter
\def\@printendtopmatter{%
  \global\let\@concepts\DASH@concepts 
  \global\let\@keywords\DASH@keywords 
  \hypersetup{pdfsubject={\@concepts}, pdfkeywords={\@keywords}}%
  \DASH@printendtopmatter}
\makeatother

\title[DASH]{ChronicleRec: Pre-training Cacheable Chronicle Tokens for Lifelong User Interest Modeling} 

\newcommand{\authorpanel}{%
  {\panelnamefont
    Chengkai Huang$^{2,*}$,
    Yubin Sheng$^{1,*}$\textsuperscript{\Letter},
    Liang Guo$^{1,*}$,
    Haoxi Liu$^{1,*}$,
    Junwei Pan$^1$,
    Shangyu Zhang$^1$,
    Zhixiang Feng$^1$,
    Chao Zhou$^1$,
    Chengguo Yin$^1$,
    Lina Yao$^2$, 
    Haijie Gu$^1$,
    Jie Jiang$^1$\par}
  \smallskip

  {\panelaffilfont\raggedright
    \mbox{$^1$Tencent Inc., China}\quad
    \mbox{$^2$University of New South Wales, Australia}\par}

{\panelmailfont\raggedright
  \mbox{$^1$yubinsheng@tencent.com, mattguo@tencent.com, haoxiliu@tencent.com}\quad
  \mbox{$^1$jonaspan@tencent.com, vitosyzhang@tencent.com, lionelfeng@tencent.com}\quad
  \mbox{$^1$derekczhou@tencent.com, turingyin@tencent.com, jerrickgu@tencent.com}\quad
  \mbox{$^2$chengkay.huang@gmail.com, $^2$lina.yao@unsw.edu.au, $^1$zeus@tencent.com}\par}%
}

\newcommand{\authornotepanel}{%
  $^*$\,Equal contribution. \quad
  \textsuperscript{\Letter}\,Corresponding author.}

\renewcommand{\shortauthors}{Huang et al.} 

\makeatletter 
\gdef\authors{Chengkai Huang, Yubin Sheng, Liang Guo, Haoxi Liu, Junwei Pan, Shangyu Zhang, Zhixiang Feng, Chao Zhou, Chengguo Yin, Haijie Gu, Lina Yao}
\makeatother

\begin{document}

\title{ChronicleRec: Pre-training Temporally Anchored Tokens for Lifelong User Modeling}

\renewcommand{\shortauthors}{Chengkai et al.}

\begin{abstract}
Modeling ultra-long user behavior sequences is crucial for industrial recommendation and online advertising, yet directly feeding thousands of historical actions into ranking models is computationally prohibitive, while truncating histories discards valuable long-range signals. Existing lifelong-interest methods typically retrieve target-relevant behaviors for each candidate, coupling long-sequence modeling with candidate scoring and incurring repeated online cost. Recent target-independent compression methods enable cached user summaries, but often append query tokens at the sequence end and use bidirectional encoding, producing unordered and potentially redundant summaries that overlook nonuniform temporal structure.
We propose ChronicleRec, a pre-train-and-transfer framework that compresses an ultra-long behavior sequence once into a small, chronologically ordered set of learnable Chronicle Tokens. ChronicleRec first applies a recency-aware multi-granularity merge, preserving fine-grained recent behaviors while coarsening distant history. It then interleaves learnable query tokens with the merged sequence and uses a lightweight causal encoder, so each query summarizes only the history before its temporal anchor. A multi-horizon design further masks different recent-history windows across parallel branches to learn complementary long-range interests. The compressor is pre-trained with a mask-and-predict objective that reconstructs held-out recent behaviors from compressed older history, aligning historical signals with near-present intent. Since Chronicle Tokens are target-independent, they can be cached per user, decoupling ultra-long sequence modeling from online candidate scoring. Experiments on KuaiRand and Tencent AdLive show that ChronicleRec consistently outperforms recent-window and single-pass compression baselines while approaching full-attention performance. Token-level analyses reveal more temporally organized and complementary representations, and a seven-day online A/B test confirms significant production gains.
\end{abstract}


\begin{teaserfigure}
 \includegraphics[width=\textwidth]{sampleteaser}
  \caption{Seattle Mariners at Spring Training, 2010.}
  \Description{Enjoying the baseball game from the third-base
  seats. Ichiro Suzuki preparing to bat.}
  \label{fig:teaser}
\end{teaserfigure}



\maketitle

\section{Introduction}

User behavior sequences are among the most informative signals in modern
recommender systems and online advertising
platforms~\cite{zhang2019deep,huang2025towards}. As users interact with a
platform over months or years, their behavior histories grow to thousands or
even tens of thousands of actions, and this \emph{lifelong} history encodes
stable long-term preferences that a short recent window cannot capture.
Empirically, extending the modeled history from tens to thousands of behaviors
yields consistent gains in click-through rate (CTR) and conversion rate (CVR)
prediction. Exploiting such ultra-long sequences at industrial scale, however, is constrained by a fundamental tension between expressiveness and efficiency.

On one end, feeding the entire history into a self-attention ranker is expressive but scales quadratically with sequence length, which is intractable when a single request must score hundreds of candidates under a strict latency
budget. On the other end, truncating the history to the most recent $L$
behaviors is cheap but silently discards the long-range interest signal that is
often decisive for conversion. The dominant industrial compromise is
\emph{target-attention retrieval}, which retrieves a small subset of
target-relevant behaviors from the long history for each candidate and scores
only those (e.g., SIM~\cite{pi2020search}, ETA~\cite{chen2021end},
SDIM~\cite{cao2022sampling}, TWIN~\cite{chang2023twin}). While effective, this
paradigm has two structural drawbacks. First, the retrieved representation is
\emph{target-dependent}: it must be recomputed for every candidate, so the cost
of long-sequence modeling is multiplied by the number of candidates per request. Second, hard retrieval by category or embedding similarity over-focuses on
behaviors superficially similar to the target, under-representing the broad, multi-scale interest structure of the user.

\begin{figure}[t]
    \centering
    \includegraphics[width=\linewidth]{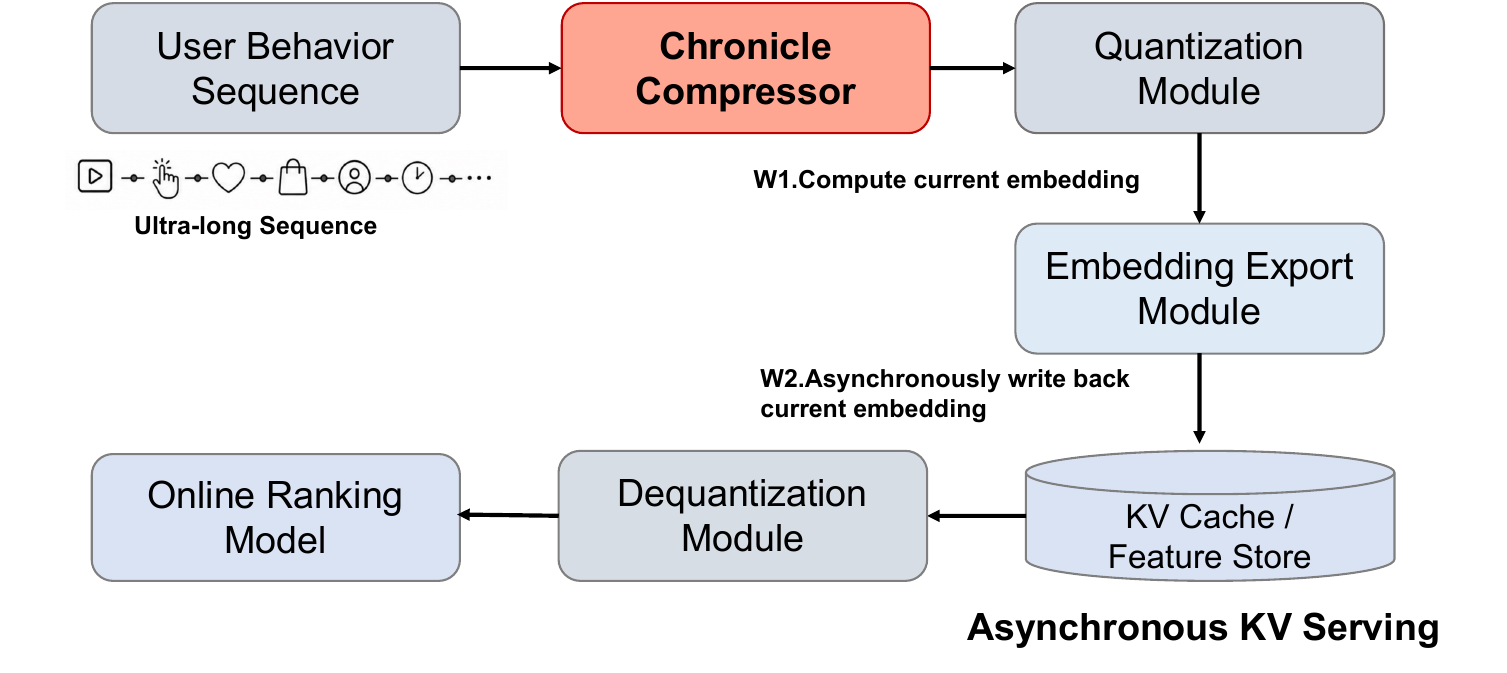}
    \caption{Asynchronous serving of Chronicle Tokens. Cached historical representations are retrieved and aggregated for downstream ranking, while the current user representation is asynchronously computed and written back to the KV cache, decoupling long-sequence encoding from online candidate scoring.}
    \label{fig:kv_serving}
\end{figure}

A recent alternative sidesteps per-candidate retrieval by \emph{compressing} the
history once into a target-independent summary that can be cached and reused
across candidates (Figure~\ref{fig:kv_serving}). The representative method,
VISTA~\cite{chen2026vista}, appends a set of learnable query tokens at the
\emph{end} of the history and lets them attend to it \emph{bidirectionally},
producing a bag of summary tokens. This is a genuine step toward
target-independent modeling, but its design leaves value on the table in two
ways. First, placing all query tokens at the tail and reading them out as an
unordered ``bag'' collapses the chronological structure of the history---the
summary is effectively \emph{sequence-in, tokens-out}. Second, bidirectional
attention lets every query peek across the entire timeline, so no query is tied
to a specific point in time and the resulting tokens tend to be redundant rather
than complementary. As our analysis later confirms, both choices are suboptimal:
tokens should be distributed \emph{along} the timeline and summarized
\emph{causally}.


We take a different stance. Rather than treating lifelong histories as
candidate-specific evidence to be retrieved at serving time, we view them as a
source for pre-training target-independent user representations. ChronicleRec
pre-trains a small and fixed-size set of temporally anchored tokens, termed
\emph{Chronicle Tokens}, from each user's lifelong behavior history. Rather than
appending the tokens at the end, we \emph{interleave} them among the history and
summarize each one \emph{causally}, so that the tokens are laid out along the
temporal axis and each token summarizes only the history up to its own anchor
position. In this sense, ChronicleRec follows a \emph{sequence-in,
sequence-out} principle: the pre-trained user representation is itself a short,
chronologically ordered token sequence, not an order-agnostic bag of latent
summaries.

Because Chronicle Tokens are target-independent, they can be computed once,
cached per user, and transferred to downstream rankers for efficient candidate
scoring. This view turns lifelong user modeling into a pre-train-and-transfer
problem: how can we learn compact, temporally organized user tokens from
long-term behavior histories while preserving the interest signals that
downstream ranking models actually need?

Realizing this view raises four challenges---the first three concern \emph{how
to compress}, and the last concerns \emph{how the compressed past should inform
the present}. \textbf{(C1) What to compress at
what granularity.} User interest is not temporally uniform: recent behaviors
are fine-grained and highly predictive, whereas distant behaviors matter mostly
as coarse, aggregated preferences. A uniform compression rate either wastes
capacity on the distant past or blurs the informative recent history.
\textbf{(C2) How to compress without leaking the future.} The tokens should
summarize history in a causal, order-aware fashion so that each token has a
well-defined temporal receptive field, rather than collapsing the sequence into
an order-agnostic pooled vector. \textbf{(C3) How to cover multiple temporal
horizons.} A single compression pass tends to be dominated by the most recent
behaviors, so distant but decisive interests are easily washed out.
\textbf{(C4) How to make the past actually inform the present.} The compressed
history summarizes the \emph{past}, but ranking concerns the \emph{present}; the
influence of past behavior on the current decision is indirect and
non-stationary, so a raw historical summary cannot be fed to the ranker as-is
and must first be aligned to the near-present interest it is meant to inform.

We address these challenges with \textbf{ChronicleRec}. To handle (C1), a
\emph{recency-aware multi-granularity merge} partitions the history into
near/mid/far segments and pools each segment at an increasing stride, producing
a chronologically ordered sequence of merged tokens that is dense near the
present and coarse in the distant past. To handle (C2), we interleave learnable
\emph{query tokens} among the merged tokens and apply a lightweight
\emph{causal} encoder, so that each query token summarizes exactly the history
up to its anchor position, with query positions placed densely near the present
and sparsely in the past. To handle (C3), a \emph{multi-branch} design runs
several compression branches in parallel, each masking out a different length
of recent history so that different branches specialize in different temporal
horizons; the branches are jointly supervised and their compact prefixes are
concatenated for the downstream ranker. Finally, to handle (C4), a mandatory
\emph{Chronicle Alignment} stage pre-trains the compressor with a
mask-and-predict objective: the most recent behaviors are held out and must be
reconstructed from the compressed older history alone, forcing the Chronicle
Tokens to project the past into a representation that is predictive of the
present. The resulting Chronicle Tokens form a short prefix that can be
prepended to any sequence ranker, either alone (replacing the long history) or
in a hybrid manner alongside a short window of fine-grained recent behaviors.


We validate ChronicleRec through a controlled study whose central claim is that, under an identical downstream ranking head, a small set of pre-trained Chronicle Tokens can recover most of the performance of full-history attention while being substantially cheaper and cacheable for online serving. Our contributions are summarized as follows:

  

\begin{itemize} 
\item We formulate lifelong user modeling as target-independent representation pre-training, and introduce temporally anchored \emph{Chronicle Tokens} that preserve the chronological structure of long-term user interests while decoupling history encoding from candidate-specific scoring. \item We propose ChronicleRec, a pre-train-and-transfer framework that combines recency-aware multi-granularity merging, causal query-token interleaving, multi-horizon compression, and alignment-oriented pre-training to learn compact, cacheable user tokens from lifelong histories. \item Extensive offline and online experiments on public and industrial datasets demonstrate that the pre-trained Chronicle Tokens improve downstream ranking, recover most of the full-attention performance with much lower serving cost, and deliver significant business gains in production. 
\end{itemize}

\section{Related Work}

\subsection{Lifelong User Interest Modeling}

Modeling long-term user behavior has evolved from memory-network approaches such as MIMN~\cite{pi2019practice}, which maintains a fixed-size user memory updated online, toward two-stage \emph{search-based} methods that dominate current industrial practice. SIM~\cite{pi2020search} introduces a General Search Unit that retrieves target-relevant behaviors from a lifelong history before an Exact Search Unit performs precise attention over the retrieved subset. Subsequent work reduces the retrieval and attention cost or improves retrieval fidelity through locality-sensitive hashing, hashing-based sampling, and consistency between the two stages (e.g., ETA, SDIM, and TWIN-style designs). Attention-based interest extractors such as DIN~\cite{zhou2018deep} and DIEN~\cite{zhou2019deep} form the backbone of the scoring stage by activating history with respect to the target item. A common thread across these methods is that the long-sequence representation is \emph{target-dependent}: relevant behaviors are selected per candidate, so serving cost grows with the number of candidates. ChronicleRec differs by
producing a \emph{target-independent} summary that is computed once per user and shared across candidates.

\subsection{Compression-Based Paradigms for Ultra-Long Sequence Modeling}

A complementary line of work compresses long inputs into a small number of
latent tokens rather than retrieving from them. In vision-language modeling,
the Q-Former of BLIP-2~\cite{li2023blip2} uses a fixed set of learnable query
tokens to distill a variable-length feature map into a compact set of queries;
Perceiver-style architectures similarly map long inputs onto a small latent
array via cross-attention. Efficient-attention techniques, including linear
attention~\cite{katharopoulos2020transformers} and other kernelized or
low-rank approximations, reduce the quadratic cost of self-attention and make
encoding long sequences tractable. In recommendation, recent efforts explore
generative and tokenized user representations, learning discrete or continuous
user codes that summarize behavior. Most closely related to our work is
VISTA~\cite{chen2026vista}, which factorizes candidate-to-history target
attention into two stages---summarizing the user history into a few hundred
tokens that are cached, and then letting candidates attend to these tokens---so
that downstream training and serving cost stays fixed as the history grows to
lifelong scale. ChronicleRec shares this target-independent, cache-and-reuse
philosophy, but differs in \emph{how} the summary is formed: VISTA performs a
single-pass summarization, whereas ChronicleRec introduces a recency-aware,
multi-granularity merge and \emph{causal} query-token interleaving so that the
compressed tokens respect the chronological, non-uniform structure of user
interest, together with a multi-branch design that explicitly covers multiple
temporal horizons. 

\section{Methodology}

\subsection{Problem Formulation and Overview}

Consider a user with an ultra-long behavior history $\mathcal{H}=(b_1,\dots,b_L)$ ordered from oldest to newest, where each behavior $b_i$ is described by a tuple of categorical fields (in our setting, the item/video ID, the author/advertiser ID, and a content tag). An embedding layer maps each behavior to a $d$-dimensional vector, yielding a sequence embedding $\mathbf{E}\in\mathbb{R}^{L\times d}$ together with a validity mask $\mathbf{m}\in\{0,1\}^{L}$ that marks padded positions (histories are left-padded, so larger indices are more recent). Given a candidate item with embedding $\mathbf{t}\in\mathbb{R}^{d}$, the ranker predicts an interaction
probability $\hat{y}=f(\mathcal{H},\mathbf{t})$, trained with binary cross-entropy against the observed label $y$ (e.g., an effective play or a conversion).

\begin{figure*}[t]
  \centering
  \includegraphics[width=\textwidth]{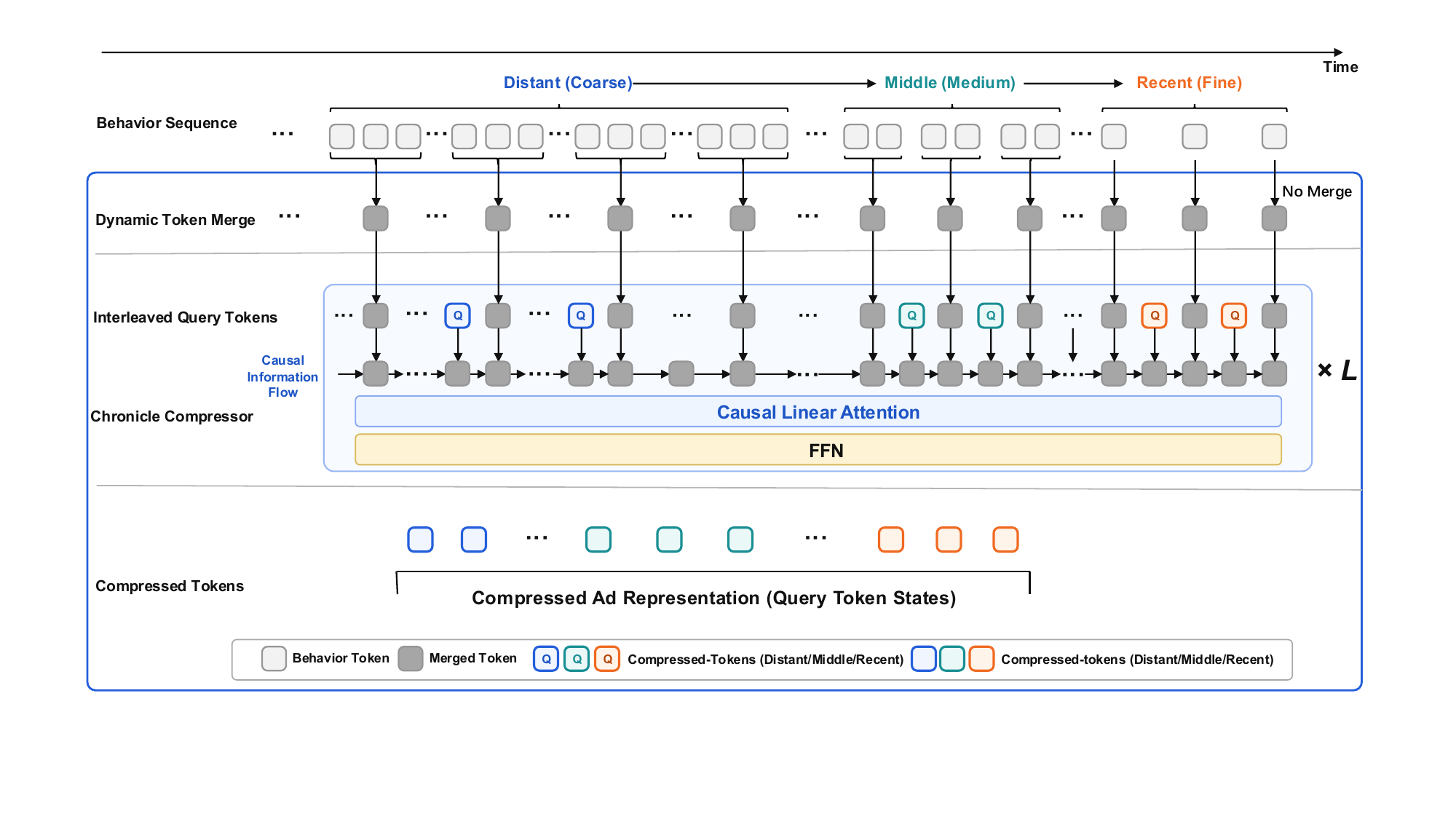}
\vspace{-2cm}
\caption{Architecture of the Chronicle Compressor. A recency-aware dynamic token merge preserves fine-grained recent behaviors while progressively compressing older history. Learnable query tokens are interleaved with the merged sequence and propagated through a causal encoder, where each query summarizes only its preceding history. The encoded query states are extracted as Chronicle Tokens, forming a compact multi-scale representation of the ultra-long behavior sequence.}
\label{fig:overview}
\end{figure*}

The core of ChronicleRec is a \emph{target-independent} compressor $g:\ (\mathbf{E},\mathbf{m})\mapsto \mathbf{C}\in\mathbb{R}^{P\times d}$ that turns the length-$L$ history into a small set of $P\!\ll\!L$ \emph{Chronicle
Tokens}. The tokens are then consumed by a shared ranking head. Because $g$ does not depend on $\mathbf{t}$, $\mathbf{C}$ is computed once per user and reused across all candidates. Figure~\ref{fig:overview} illustrates the pipeline; the compressor consists of three components described below: a recency-aware multi-granularity merge (\S\ref{sec:merge}), causal query-token interleaving (\S\ref{sec:query}), and a multi-branch multi-horizon design (\S\ref{sec:multi}).


\subsection{Recency-Aware Multi-Granularity Merge}
\label{sec:merge}

Because recent behaviors are more predictive than distant ones (C1), we compress the history non-uniformly. We split the effective history into three contiguous segments ordered from old to new: a \emph{far} segment (the distant past), a \emph{mid} segment, and a \emph{near} segment (the most recent behaviors). Given a budget of $n$ preserved recent behaviors and a target mid length, the segment boundaries are:
\begin{equation}
  \text{near} = \text{last } n \text{ behaviors},\quad
  \text{mid},\ \text{far} = \text{the remaining older history}.
\end{equation}
Each segment is summarized by a masked average pooling over a sliding window with a segment-specific kernel and stride. The near segment is kept at full resolution (stride $1$, i.e., no merging); the mid segment uses a moderate kernel/stride; and the far segment uses a large kernel/stride for aggressive aggregation. Formally, for a window $w$ covering source positions $\mathcal{S}_w$, the merged token is:
\begin{equation}
  \mathbf{u}_w=\frac{\sum_{i\in\mathcal{S}_w} m_i\,\mathbf{E}_i}
                    {\max\!\big(\sum_{i\in\mathcal{S}_w} m_i,\ 1\big)},
  \qquad
  \tilde m_w=\mathbb{1}\!\Big[\textstyle\sum_{i\in\mathcal{S}_w} m_i>0\Big],
\end{equation}
where the mask $m_i$ ensures padded positions never leak into a merged token. Concatenating the windows across the three segments produces a chronologically ordered merged sequence $\mathbf{U}\in\mathbb{R}^{M\times d}$ with mask $\tilde{\mathbf{m}}$, where $M\!\ll\!L$ and the resolution is dense near the present and coarse in the distant past. For short-history users, the far windows fall on left padding and are masked out, so the merge equivalently compresses only the behaviors that actually exist. The window-to-source assignment is static given the maximum length, so the merge is implemented as a single vectorized gather-and-pool operation with negligible overhead.

\subsection{Causal Query-Token Interleaving}
\label{sec:query}

To obtain a fixed-size, order-aware summary (C2), we introduce $P$ learnable \emph{query tokens} $\mathbf{Q}\in\mathbb{R}^{P\times d}$ and interleave them among the merged tokens $\mathbf{U}$ according to a set of anchor positions. The anchors are placed by their offset from the most recent end of the merged sequence, densely near the present and increasingly sparse toward the past, so that recent history is summarized at finer temporal resolution. Interleaving yields a mixed sequence in which each query token is anchored at a specific chronological position.

The mixed sequence is processed by a lightweight Transformer encoder with a \emph{causal} attention mask: every position attends only to positions at or before it, and padded merged tokens are additionally masked out as keys. As a result, each query token summarizes exactly the history up to its anchor, giving it a well-defined temporal receptive field. To avoid degenerate all-masked rows (which arise for the distant queries of short-history users and would otherwise produce numerical instabilities in the attention softmax), the self-position is always kept visible. After encoding, we read out the encoder states at the query positions as the compressed output:
\begin{equation}
\mathbf{C}=\mathrm{Encoder}_{\text{causal}}\big([\mathbf{U};\mathbf{Q}]_{\text{interleaved}}\big)\big|_{\text{query positions}}
  \ \in\ \mathbb{R}^{P\times d}.
\end{equation}
These are the Chronicle Tokens. We also consider a Q-Former–style variant that first down-samples the history with a strided convolutional sub-sequence encoder and then lets a set of query tokens attend to it through a bidirectional linear attention encoder~\cite{katharopoulos2020transformers}; we treat this variant as an alternative instantiation of the compressor and compare against it in experiments.

\begin{figure}[t]
  \centering
  \includegraphics[width=0.5\textwidth]{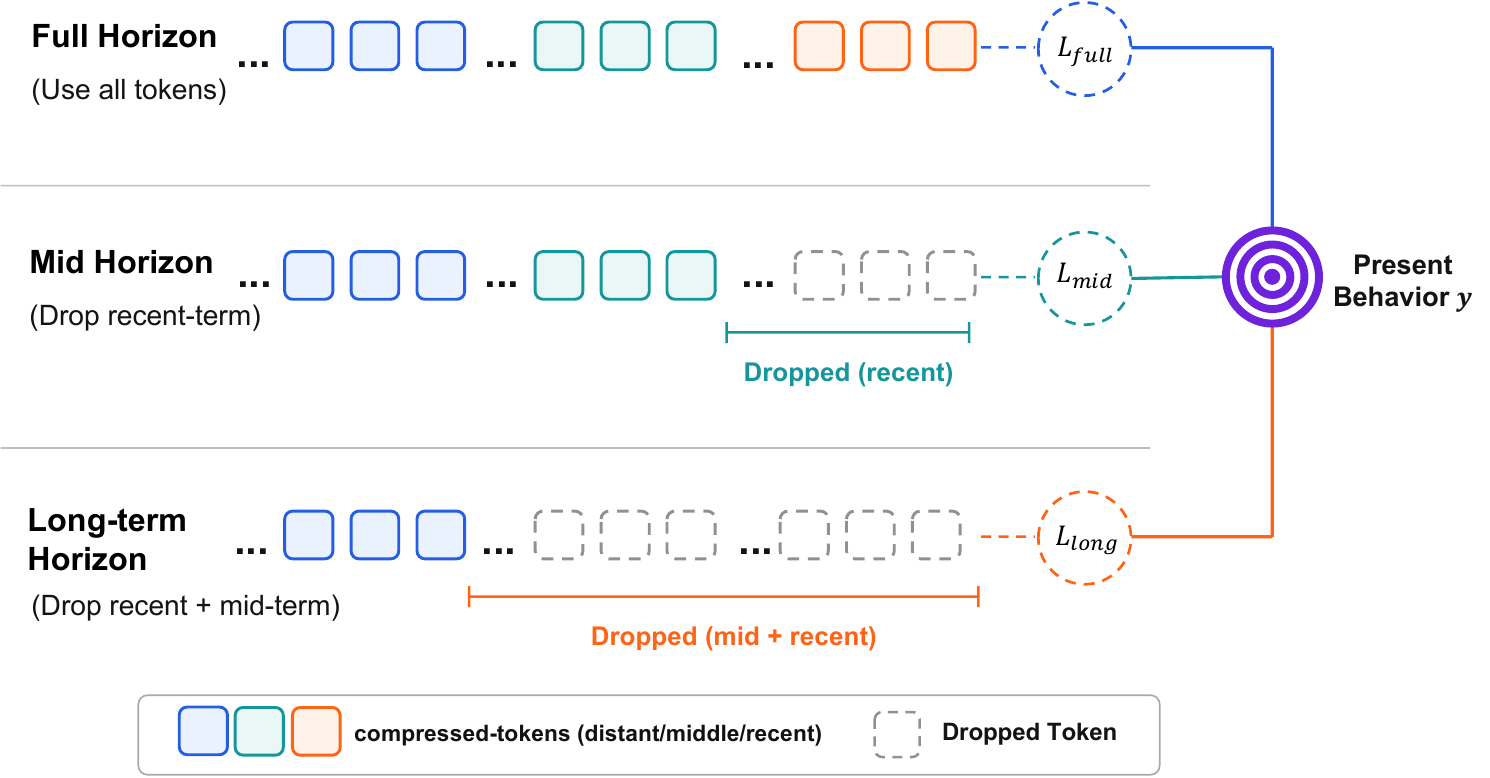}
\caption{
Multi-branch multi-horizon learning. Different branches mask progressively larger recent-history windows before compression, forcing them to model increasingly distant temporal horizons. Each branch is independently supervised to predict the present behavior, encouraging complementary interest representations across multiple temporal scales.
}\label{fig:details}
\end{figure}

\subsection{Multi-Branch Multi-Horizon Compression}
\label{sec:multi}

A single compression pass is dominated by the most recent behaviors, so distant but decisive interests can be washed out (C3). We therefore employ $B$ parallel compression branches. Branch $j$ masks out the most recent $\delta_j$ behaviors before compression, with $0=\delta_1<\delta_2<\dots<\delta_B$, so that branches with larger $\delta_j$ are forced to attend to progressively more distant horizons. Each branch has its own query tokens and encoder, and produces a prefix $\mathbf{C}^{(j)}\in\mathbb{R}^{P\times d}$. The branch prefixes are scaled by per-branch weights $w_j$ and concatenated into the final Chronicle token set:
\begin{equation}
\mathbf{C}=\big[\,w_1\mathbf{C}^{(1)};\,w_2\mathbf{C}^{(2)};\,\dots;\,w_B\mathbf{C}^{(B)}\,\big]
  \ \in\ \mathbb{R}^{(B\cdot P)\times d}.
\end{equation}


\paragraph{Chronicle Alignment.}
Although the Chronicle Tokens summarize the historical behavior sequence, the
downstream ranker ultimately needs a representation that is predictive of the
near-present user intent. We therefore introduce a Chronicle Alignment objective
to align the compressed past with the present decision signal. Specifically, for
each horizon branch $j$, we mask out a recent behavior window before compression
and feed the resulting branch prefix $\mathbf{C}^{(j)}$ into a lightweight
target-attention head to predict the held-out target label $\hat y_j$. The
branches are jointly optimized with a weighted alignment loss:
\begin{equation}
  \mathcal{L}_{\text{align}}
  =
  \sum_{j=1}^{B}
  \bar w_j\,
  \mathrm{BCE}(\hat y_j, y),
  \qquad
  \bar w_j
  =
  \frac{w_j}{\sum_{k=1}^{B} w_k}.
\end{equation}
This objective forces each branch to project its compressed historical context
toward the near-present interest signal, while encouraging different horizons to
retain complementary predictive information. At inference time, the auxiliary
heads are removed, and the concatenated prefix $\mathbf{C}$ is passed to the
shared ranker as a compact Chronicle memory.

\subsection{Downstream Ranking and Training}
\label{sec:rank}

The Chronicle Tokens are consumed by a shared ranking head that stacks a few Pre-LN self-attention layers followed by a target-attention read-out, in which the candidate embedding $\mathbf{t}$ acts as the query over the tokens to obtain an interest vector, which is concatenated with $\mathbf{t}$ and scored by an MLP. ChronicleRec supports two usage modes: a \emph{compress-only} mode, in which the head consumes the Chronicle Tokens alone (replacing the raw long history), and a
\emph{hybrid} mode, in which the tokens are prepended as a prefix to a short window of fine-grained recent behaviors before the head. Crucially, all variants share the same head so that any difference in accuracy is attributable to the compression step rather than to head capacity. The compressor is first pre-trained with the Chronicle Alignment objective (Stage 1) and then loaded into this ranker for end-to-end training (Stage 2), with the compressor either frozen or fine-tuned.

\section{Experiment}


\subsection{Experimental Setup}

\paragraph{Datasets.}
We use two datasets summarized in Table~\ref{tab:datasets}. \textbf{KuaiRand} (the KuaiRand-27K split)~\cite{gao2022kuairand} is a public short-video recommendation benchmark with exceptionally long behavior histories---on average close to $12$K events per user and up to $228$K for the most active users---over an extremely sparse video space, which makes it an ideal testbed for ultra-long sequence modeling; we use the click signal ($\texttt{is\_click}$) as the positive label. \textbf{Tencent AdLive} is an industrial online-advertising (live-streaming) dataset sampled from a production system, in which nearly $200$K users are sampled from a pool of over $12.7$M, with an average history length close to $1{,}070$ behaviors and a click
as the positive label. Because raw histories far exceed what a ranker can consume, they are left-padded and truncated to a maximum length $L_{\max}$ ($2{,}048$ for KuaiRand, $4{,}000$ for AdLive); each behavior is represented by its item/creative, author/advertiser, and tag/industry IDs.

\begin{table}[t]
  \centering
  \caption{Statistics of the two datasets. ``Hist.\ len.'' refers to the raw per-user history length before truncation; $L_{\max}$ is the truncation cap used during training. ``$-$'' denotes a statistic not applicable or not disclosed for the public split.}
  \label{tab:datasets}
  \small
  \setlength{\tabcolsep}{4pt}
  \begin{tabular}{lrr}
    \toprule
    Statistic & KuaiRand-27K & Tencent AdLive \\
    \midrule
    \#Users            & 27{,}285      & 199{,}914 \\
    \#Items            & $\sim$32M     & 10{,}181{,}793 \\
    \#Tags/Industries  & 59            & 317 \\
    \#Interactions     & $\sim$322M    & 213{,}859{,}057 \\
    Avg.\ hist.\ len.  & 11{,}811.6    & 1{,}069.8 \\
    \bottomrule
  \end{tabular}
\end{table}

\paragraph{Baselines and variants.}
To isolate the effect of compression, all methods share the \emph{same} ranking
head and differ only in how the ultra-long history is turned into tokens:
\begin{itemize}
  \item \textbf{Full-Attn}: full self-attention over the entire history---the
  expensive but expressive upper reference.
  \item \textbf{Short-Attn}: self-attention over only the most recent $100$
  behaviors---the cheap recent-window baseline.
  \item \textbf{VISTA}~\cite{chen2026vista}: a two-stage
  compression-based long-sequence encoder that summarizes the user history into
  a few hundred cacheable tokens which candidates then attend to; we adopt it as
  a representative single-pass compression baseline.
  \item \textbf{ChronicleRec}: our recency-aware three-segment compressor
  (\S\ref{sec:merge}--\ref{sec:query}), consuming the Chronicle Tokens alone.
  \item \textbf{Hybrid-ChronicleRec}: the Chronicle Tokens prepended as a prefix to the recent-window tokens before the shared head (\S\ref{sec:rank}).
  \item \textbf{Multi-ChronicleRec}: the full multi-branch multi-horizon design
  with deep supervision (\S\ref{sec:multi}).
\end{itemize}

\paragraph{Metrics.}
We report GAUC (grouped by user) as the primary accuracy metric. All models share the embedding layer, head architecture, and training schedule, so that any accuracy difference is attributable to how the history is turned into tokens rather than to head capacity.

\paragraph{Architecture.}
We set the model dimension $d_{\text{model}}{=}64$ (video/author/tag embeddings of $64/16/16$). The chronicle compressor uses a recency-aware multi-scale merge: the most recent $\text{near\_keep}{=}100$ events are kept $1{:}1$, a mid segment of length $400$ is merged with kernel/stride=$8/4$, and the far segment with kernel/stride=$10/10$; a causal Transformer encoder ($4$ layers, $4$ heads, dropout $0.1$) with $P{=}11$ query tokens produces the chronicle tokens by default. The multi-branch variant uses drop offsets $\{\delta_j\}{=}\{0,100,300,600,1000\}$ with branch weights $\{w_j\}{=}\{1.0,0.5,0.4,0.3,0.2\}$. The prediction head is a $2$-layer target-aware attention module ($4$ heads, dropout $0.1$).

\paragraph{Training.}
We optimize a BCE loss with Adam (lr=$10^{-3}$, weight decay=$10^{-6}$), using linear warmup followed by optional cosine decay. The batch size is $256$ (KuaiRand) / $512$ (AdLive),
models are trained for $3$ epochs with early stopping.

\subsection{Main Results}

Table~\ref{tab:main} reports the main comparison on both datasets, measured by GAUC. Several observations stand out. First, extending the modeled history clearly helps: Full-Attn over the entire history reaches $0.5601$ GAUC on KuaiRand and $0.8071$ on AdLive, substantially above the recent-window
Short-Attn ($0.5307$ and $0.7950$), confirming that the distant history carries a decisive interest signal that a short window discards. Second, and central to our claim, a small set of Chronicle Tokens recovers most of this gap at a fraction of the cost: multi-ChronicleRec attains $0.5580$ GAUC on KuaiRand, within $0.0021$ of Full-Attn while feeding the head only a compact prefix rather than the full sequence. Third, ChronicleRec compares favorably to the single-pass compression baseline VISTA ($0.5518$): the recency-aware three-segment compressor already improves to $0.5536$, the hybrid usage to
$0.5541$, and the multi-branch design to $0.5580$, indicating that both the chronological, non-uniform compression and the multi-horizon design contribute beyond generic latent compression. The same ordering holds on the industrial AdLive dataset: VISTA reaches $0.7995$ GAUC, the three-segment ChronicleRec $0.8010$, Hybrid-ChronicleRec $0.8030$, and Multi-ChronicleRec $0.8034$, all well above Short-Attn's $0.7950$ and closely approaching Full-Attn ($0.8071$). 
Overall, these results show that, under an identical prediction head, a small number of Chronicle Tokens compressed from an ultra-long history can match or approach full attention while clearly outperforming recent-window baselines.

\begin{table}[t]
  \centering
  \caption{
    Main results (GAUC) on KuaiRand and Tencent AdLive.
    All methods share the same ranking head; the only variable is how the
    ultra-long history is turned into tokens. Best results are in
    \textbf{bold}, and second-best results are \underline{underlined}.
  }
  \label{tab:main}
  \small
  \setlength{\tabcolsep}{3pt}
  \begin{tabularx}{\linewidth}{@{}lXcc@{}}
    \toprule
    Method & Mechanism & KuaiRand & AdLive \\
    \midrule
    Full-Attn
      & Full self-attention
      & \textbf{0.5601}
      & \textbf{0.8071} \\
    Short-Attn
      & Recent-window attention
      & 0.5307
      & 0.7950 \\
    VISTA
      & Single-pass compression
      & 0.5518
      & 0.7995 \\
    \midrule
    ChronicleRec
      & Three-segment compression
      & 0.5536
      & 0.8010 \\
    Hybrid-ChronicleRec
      & Recent + far compression
      & 0.5541
      & 0.8030 \\
    Multi-ChronicleRec
      & Multi-horizon compression
      & \underline{0.5580}
      & \underline{0.8034} \\
    \bottomrule
  \end{tabularx}
\end{table}

\subsection{Query-Token Placement and Attention Direction}
\label{sec:placement}

A central design choice in ChronicleRec is how learnable query tokens are positioned relative to the merged history and how information is propagated among them (\S\ref{sec:query}). We study two factors: query placement, comparing \emph{interleaved} queries distributed along the merged sequence with \emph{end} placement that appends all queries after the history, and attention direction, comparing \emph{causal} and \emph{bidirectional} encoding. Table~\ref{tab:placement} reports the resulting GAUC on Tencent AdLive.

The interleaved-causal configuration used by ChronicleRec achieves the best performance, reaching $0.8034$ GAUC, compared with $0.7890$ for end-causal, $0.7930$ for interleaved-bidirectional, and $0.7951$ for end-bidirectional. Notably, under causal encoding, distributing queries along the sequence improves GAUC by $0.0144$ over placing them all at the end. Likewise, for interleaved queries, causal encoding improves GAUC by $0.0104$ over bidirectional encoding.

These results highlight the complementary roles of query placement and information flow. Interleaving assigns queries to different locations along the user timeline, while causal encoding turns these locations into well-defined temporal anchors by restricting each query to its preceding history. Their combination therefore produces a chronologically structured set of summaries rather than multiple queries that repeatedly summarize the same global sequence. This empirical advantage supports the interleaved query-token design at the core of ChronicleRec.

\begin{table}[t]
  \centering
  \caption{Query-token placement and attention direction on Tencent AdLive (best GAUC). ``Interleaved'' anchors queries among the merged tokens; ``End'' appends them at the sequence end. The interleaved-causal design used by ChronicleRec ranks first. Best in \textbf{bold}.}
  \label{tab:placement}
  \small
  \begin{tabular}{llc}
    \toprule
    Placement & Direction & GAUC \\
    \midrule
    Interleaved & Causal        & \textbf{0.8034} \\
    End         & Causal        & 0.7890 \\
    Interleaved & Bidirectional & 0.7930 \\
    End         & Bidirectional & 0.7951 \\
    \bottomrule
  \end{tabular}
\end{table}

\subsection{Ablation Study}
\label{sec:ablation}

We ablate the components of ChronicleRec in Table~\ref{tab:ablation},
reporting GAUC on Tencent AdLive: (a) replacing the recency-aware
multi-granularity merge with a uniform-stride merge (\emph{w/o recency}); (b) removing the causal mask, i.e., bidirectional interleaving (\emph{w/o causal}); (c) using a single branch instead of the multi-branch multi-horizon design (\emph{w/o multi-branch}); (d) removing deep supervision; and (e) removing the Chronicle Alignment stage, i.e., training the ranker with an unaligned compressor.

All components contribute positively to performance. Removing the
recency-aware merge decreases GAUC by $0.0076$, confirming the importance of non-uniform, chronology-aware compression. Removing the causal mask and the multi-branch design leads to drops of $0.0074$ and $0.0072$, respectively, demonstrating the benefits of temporally ordered information flow and explicit multi-horizon modeling. Deep supervision also provides a consistent improvement, with its removal reducing GAUC by $0.0039$. Finally, removing Chronicle Alignment causes the largest degradation ($-0.0086$ GAUC), highlighting the importance of aligning the compressed historical representation with near-present user interest before downstream ranking.


\begin{table}[t]
  \centering
  \caption{Ablation study on Tencent AdLive (GAUC). $\Delta$ is relative to the full model.}
  \label{tab:ablation}
  \small
  \setlength{\tabcolsep}{5pt}
  \begin{tabular}{lcc}
    \toprule
    Variant & GAUC & $\Delta$ \\
    \midrule
    multi-ChronicleRec (full)   & 0.8034 & --        \\
    \ \ w/o recency-aware merge & 0.7958 & $-0.0076$ \\
    \ \ w/o causal mask         & 0.7960 & $-0.0074$ \\
    \ \ w/o multi-branch        & 0.7962 & $-0.0072$ \\
    \ \ w/o deep supervision    & 0.7995 & $-0.0039$ \\
    \ \ w/o Chronicle Alignment & 0.7948 & $-0.0086$ \\
    \bottomrule
  \end{tabular}
\end{table}

\subsection{Long-Range Signal Analysis}
\label{sec:longrange}

To probe how much predictive signal resides in the \emph{distant} history, we mask out the most recent $N$ behaviors of every user and let ChronicleRec rely solely on the remaining older history. Table~\ref{tab:maskrec} reports the results on KuaiRand (GAUC). As expected, performance decreases monotonically as more recent behaviors are hidden, dropping from $0.5580$ with the full history to $0.5499$ once the most recent $1{,}000$ behaviors are removed, which confirms that recent behaviors are the most predictive. Crucially, however, this degradation is gradual rather than catastrophic. Even after masking the entire recent window of $1{,}000$ events, the model still attains $0.5499$ GAUC, far exceeding the recent-window Short-Attn baseline that observes only such recent behaviors ($0.5307$ GAUC, Table~\ref{tab:main}). In other words, the distant history alone already carries a strong, self-contained interest signal that ChronicleRec is able to compress and exploit. This finding directly supports the motivation behind the \emph{sequence-in, sequence-out} paradigm: modeling the long tail of user history yields incremental information well beyond the recent window, and the Chronicle Tokens faithfully preserve it.

\begin{table}[t]
  \centering
  \caption{Long-range signal analysis on KuaiRand (GAUC). We report performance when the most recent $N$ behaviors are masked out and only the older history is used; $\Delta$ is measured relative to the full history ($N{=}0$).}
  \label{tab:maskrec}
  \small
  \setlength{\tabcolsep}{6pt}
  \begin{tabular}{ccc}
    \toprule
    Masked recent $N$ & GAUC & $\Delta$ \\
    \midrule
    $0$ (full)  & 0.5580 & --       \\
    $100$       & 0.5532 & $-0.0048$ \\
    $300$       & 0.5529 & $-0.0051$ \\
    $500$       & 0.5527 & $-0.0053$ \\
    $1{,}000$   & 0.5499 & $-0.0081$ \\
    \bottomrule
  \end{tabular}
\end{table}

\subsection{Complexity and Efficiency}
\label{sec:eff}

Table~\ref{tab:efficiency} compares the computational cost of different
long-history modeling strategies. Full-Attn is substantially more expensive,
requiring $49.1$\,GB of GPU memory and nearly $3000$ minutes of training due to
self-attention over the entire behavior history. In contrast, ChronicleRec
reduces the memory footprint to $28.0$\,GB and the training time to $620.3$
minutes, corresponding to approximately a $4.8\times$ training speedup over
Full-Attn while retaining most of its predictive performance.
Hybrid-ChronicleRec further reduces memory usage to $24.8$\,GB with a similar
training cost of $614.3$ minutes, whereas Multi-ChronicleRec incurs additional
cost ($34.8$\,GB and $855.1$ minutes) in exchange for the accuracy gains brought
by its multiple temporal-horizon branches.

Importantly, these additional compression costs need not be paid for every
candidate at serving time. Since Chronicle Tokens are target-independent, they
can be computed once per user and stored in the feature cache for reuse across
candidate scoring, decoupling ultra-long sequence encoding from the online
ranking path. Parameter counts are also comparable among the long-history
methods ($\approx\!1.23$\,B), indicating that their performance differences
primarily arise from the sequence modeling strategy rather than model capacity.

\begin{table}[t]
  \centering
  \caption{Complexity and efficiency. GPU memory and total training
  time are measured under the same batch size and hardware. 
  }
  \label{tab:efficiency}
  \small
  \setlength{\tabcolsep}{4pt}
  \begin{tabular}{lrrr}
    \toprule
    Method & Params (M) & Mem (GB) & Train (min) \\
    \midrule
    Full-Attn           & 1229.9 & 49.1 & 2997.0 \\
    Short-Attn          &  661.2 & 16.2 & 1368.8 \\
    VISTA               & 1230.1 & 24.8 &  610.9 \\
    \midrule
    ChronicleRec        & 1230.0 & 28.0 &  620.3 \\
    Hybrid-ChronicleRec & 1230.1 & \textbf{24.8} & 614.3 \\
    Multi-ChronicleRec  & 1230.4 & 34.8 &  855.1 \\
    \bottomrule
  \end{tabular}
\end{table}

\subsection{Token Similarity Analysis}
\label{sec:token_similarity}

We further examine whether the compressed tokens encode complementary temporal
information or collapse into redundant summaries. To this end, we visualize the
pairwise cosine similarity between compressed token positions. Tokens are
ordered from the farthest temporal anchor to the most recent anchor and grouped
into far, middle, and recent temporal regions.

\begin{figure}[t]
    \centering
    \includegraphics[width=\linewidth]{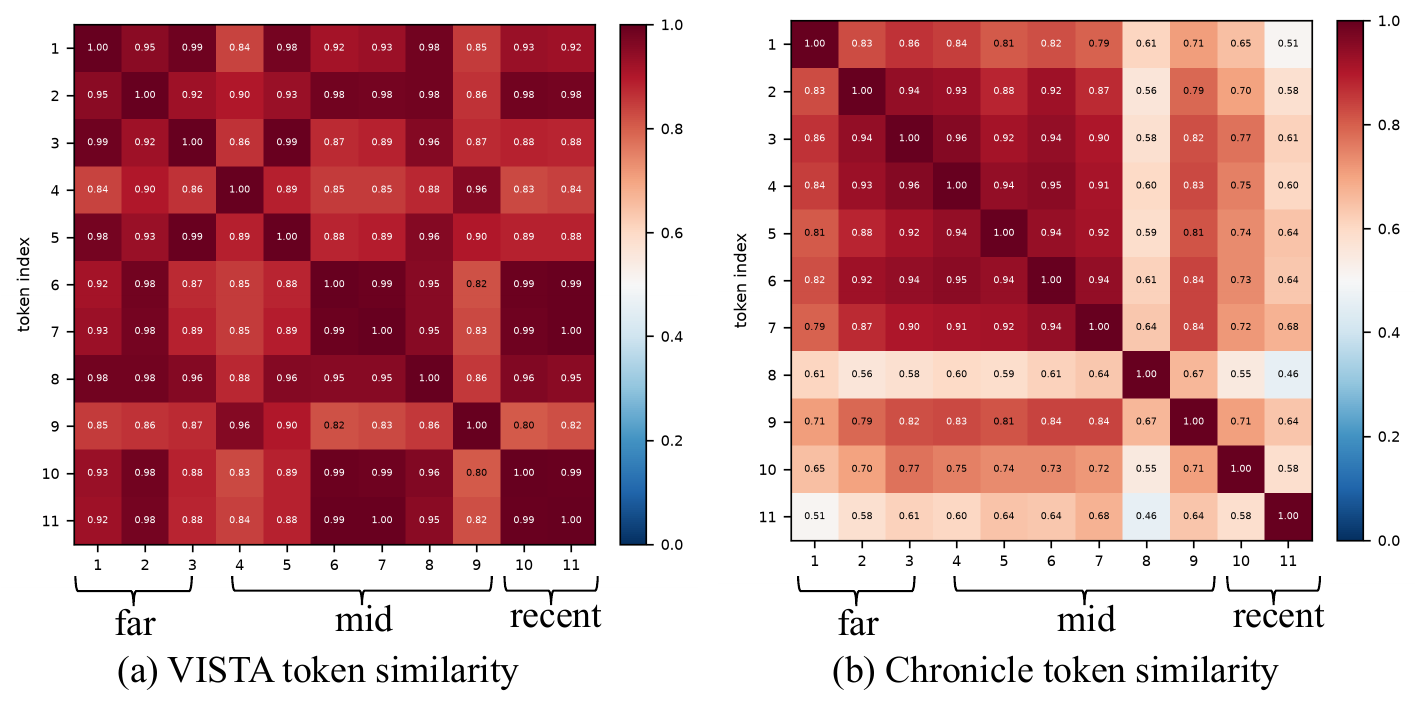}
    \caption{
    Pairwise cosine similarity between compressed token positions. Token
    indices are ordered from the farthest to the most recent temporal anchor
    and grouped into far, middle, and recent regions. VISTA exhibits uniformly
    high similarity across many non-local token pairs, suggesting substantial
    redundancy among its compressed summaries. In contrast, ChronicleRec shows
    a clearer temporal organization: tokens within nearby temporal regions
    remain strongly correlated, while cross-region similarities are relatively
    lower. This indicates that Chronicle Tokens preserve more temporally
    specialized and complementary information.
    }
    \label{fig:token_similarity}
\end{figure}

Figure~\ref{fig:token_similarity} compares the pairwise cosine similarity of
compressed tokens produced by VISTA and ChronicleRec. VISTA shows consistently
high similarity across a large portion of the token pairs, including pairs from
non-adjacent temporal regions. This suggests that end-appended bidirectional
query tokens tend to extract overlapping global summaries from the full history,
resulting in redundant token representations.

ChronicleRec, in contrast, exhibits a more structured similarity pattern. The
learned Chronicle Tokens form distinguishable temporal blocks: tokens within
the same or nearby temporal regions remain highly correlated, whereas tokens
from more distant regions have noticeably lower similarity. This organization
is consistent with interleaved causal query compression. Since each query token
is inserted at a specific temporal anchor and can only aggregate information
from its historical context, different Chronicle Tokens are encouraged to
summarize different stages of the user's behavioral trajectory.

This analysis provides qualitative evidence that ChronicleRec improves not only the amount of compressed information but also its organization across token
positions. Compared with VISTA, Chronicle Tokens exhibit clearer temporal
specialization and lower cross-region redundancy, making them more informative
prefix representations for downstream ranking.

\begin{figure}[t]
    \centering
    \includegraphics[width=\linewidth]{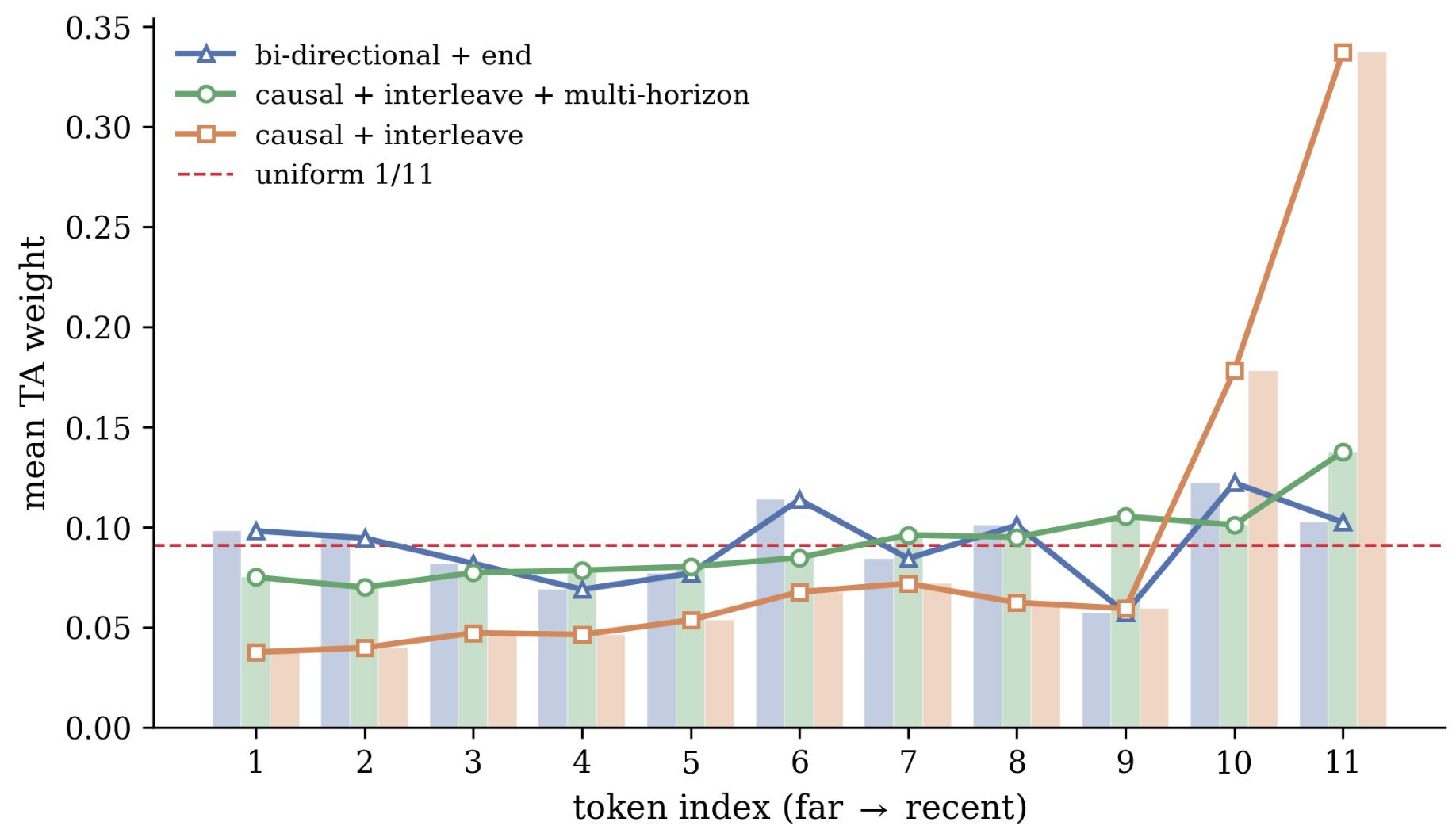}
    \caption{
    Target-attention weights over Chronicle Token positions for a representative user. Token indices range from the farthest to the most recent temporal
    anchor, and the dashed line denotes uniform attention ($1/P$). Multi-horizon ChronicleRec distributes attention across a broader range of temporal positions, whereas removing multi-horizon modeling concentrates substantially more attention on the most recent tokens.}
    \label{fig:target_attention}
\end{figure}

We next investigate whether the temporal diversity encoded by Chronicle Tokens is actually utilized by the downstream ranker.
Figure~\ref{fig:target_attention} visualizes the target-attention weights over Chronicle Token positions for a representative user. With multi-horizon modeling, attention is distributed across a broad range of temporal positions.
Although recent tokens receive somewhat larger weights, several middle- and long-term tokens remain close to or above the uniform baseline, suggesting that candidate scoring draws on information from multiple temporal horizons.

Removing multi-horizon modeling produces a markedly more concentrated distribution. In this example, the most recent token receives roughly one third of the total attention mass, while distant and middle-horizon tokens receive substantially smaller weights. This qualitative contrast suggests that multi-horizon learning prevents the compressed representation from being dominated by recent behaviors and keeps longer-range signals accessible to the downstream ranker.

Overall, the effective-rank, token-similarity, and target-attention analyses provide complementary evidence that ChronicleRec preserves both representational diversity and temporal organization after compression. Chronicle Tokens capture distinct stages of the user history, while the downstream ranker can selectively utilize information from different temporal horizons during candidate scoring.

\subsection{Cumulative Information Gain of Chronicle Tokens}
\label{sec:mi_gain}

To complement the representation-level analyses above, we further examine
whether adding compressed tokens contributes target-relevant information beyond
that already captured by the preceding token prefix. For the interleaved
variants, let $\mathbf{c}_p$ denote the compressed token associated with the
$p$-th temporal anchor, ordered from the farthest to the most recent anchor.
For the first $p$ tokens, we estimate the cumulative mutual information (MI)
with the prediction target as:
\begin{equation}
    \widetilde{I}_p
    =
    \widetilde{I}
    \left(
    y ;
    \mathbf{c}_{1:p}
    \right),
\end{equation}
where $y$ denotes the prediction target and
$\widetilde{I}(\cdot)$ is the debiased MI estimate. For exact mutual information, the incremental gain from adding the $p$-th token satisfies:
\begin{equation}
    I(y;\mathbf{c}_{1:p})
    -
    I(y;\mathbf{c}_{1:p-1})
    =
    I(y;\mathbf{c}_p \mid \mathbf{c}_{1:p-1}).
\end{equation}
So a sustained increase in cumulative MI reflects additional target-relevant
information that is not already captured by the preceding prefix. Since
$\widetilde{I}$ is estimated from finite samples, we focus on the overall
trend of the curve rather than requiring strict monotonicity.

\begin{figure}[t]
    \centering
    \includegraphics[width=\linewidth]{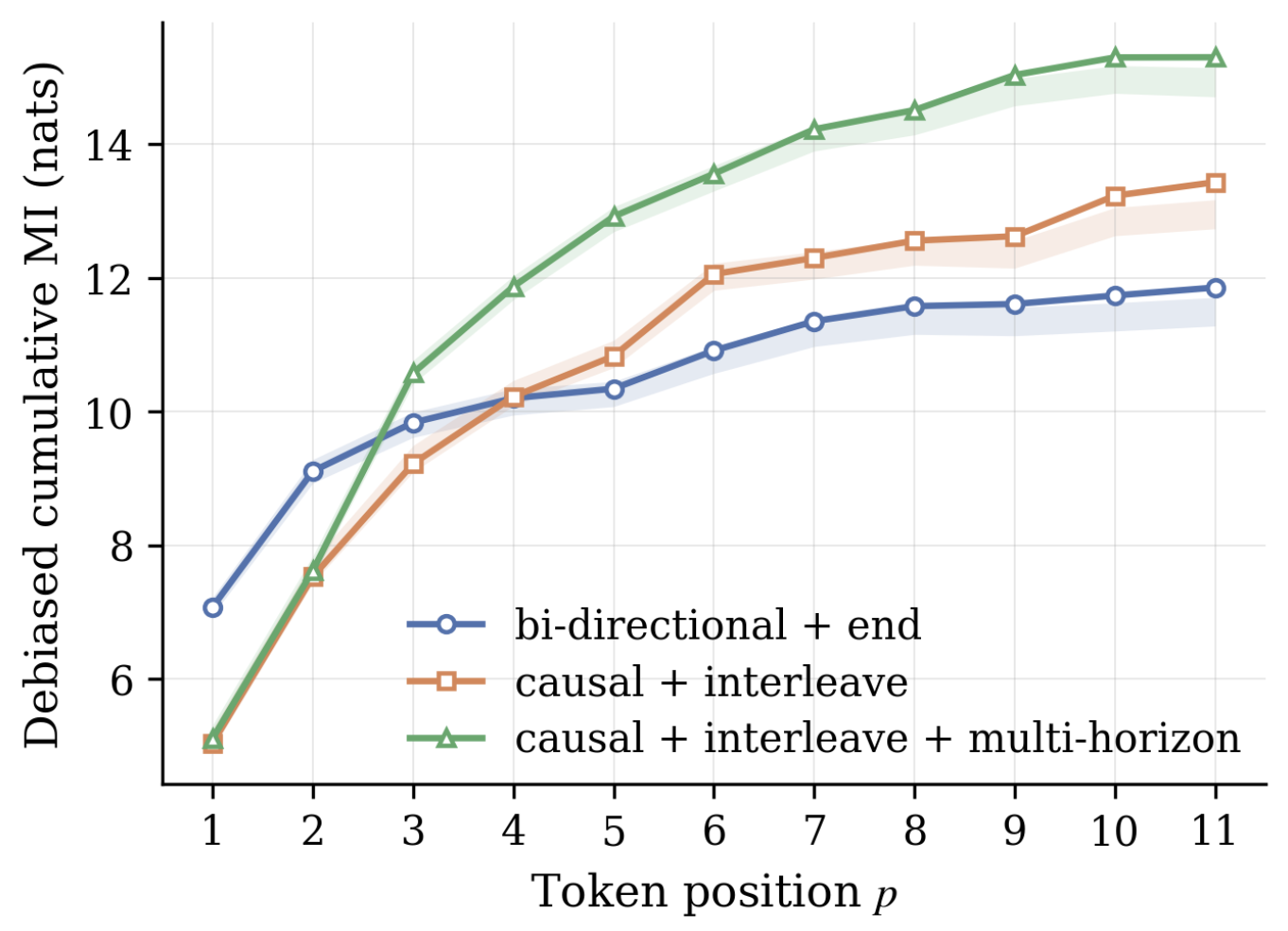}
    \caption{
    Debiased cumulative mutual information between the prediction target and
    the compressed-token prefix. For the interleaved variants, tokens are
    accumulated from the farthest to the most recent temporal anchor; the
    end-appended bidirectional baseline uses its fixed query-slot order.
    Compared with the end-appended baseline, causal interleaving yields a more
    sustained increase in target-relevant information, while multi-horizon
    learning achieves the largest cumulative MI over most of the token range.}
    \label{fig:mi_gain}
\end{figure}

Figure~\ref{fig:mi_gain} compares three token-construction strategies.
The \emph{bidirectional + end} variant appends all query tokens after the
behavior sequence and allows each query to attend to the full history. Its
cumulative MI rises rapidly with the first few query slots but grows only
modestly thereafter, suggesting substantial overlap in the target-relevant
information captured by different end-appended queries. In contrast, the
\emph{causal + interleave} design assigns queries to distinct positions along
the behavior timeline and restricts each query to its preceding context.
Its cumulative MI increases more steadily and eventually exceeds that of the
end-appended variant, consistent with the distinct temporal receptive fields
induced by causal interleaving.

The \emph{causal + interleave + multi-horizon} variant achieves the highest
cumulative MI over most of the token range. This result suggests that
multi-horizon learning further encourages the compressed representation to
retain target-relevant signals from different temporal contexts, rather than
allowing all query tokens to be dominated by the same portion of the history.
Together with the effective-rank and token-similarity analyses, these results
provide complementary evidence that ChronicleRec learns a temporally
structured representation in which additional token positions contribute
useful information for downstream prediction.

\subsection{Online A/B Test}

We deploy ChronicleRec in the pCVR prediction scenario of Weixin Moments Ads, a large-scale industrial recommendation setting that requires real-time estimation of users’ conversion probabilities for candidate ads under massive online traffic and stringent latency constraints. The production model adopts a MixFormer architecture to jointly model large-scale heterogeneous non-sequential features and user behavior sequences, where NS-Tokens are used for non-sequential feature interaction and S-Tokens for sequential behavior modeling. In our deployment, the compressed Chronicle Tokens are incorporated into both the NS-Token and S-Token pathways, enabling long-term user behavior information to be effectively integrated into the existing ranking model without altering its main architecture.
ChronicleRec increases GMV by \textbf{+1.61\%}, with
a $95\%$ confidence interval of $[0.678\%, 2.547\%]$ that excludes zero, i.e., a statistically significant improvement. Given the scale of the platform, a lift of this magnitude on GMV translates into substantial incremental revenue, demonstrating that the offline gains transfer to real business value under production latency constraints.


\section{Conclusion}

We introduced ChronicleRec, a pre-train-and-transfer framework for lifelong user modeling. ChronicleRec learns temporally anchored Chronicle Tokens from long-term user behavior histories through recency-aware merging, causal query-token interleaving, multi-horizon compression, and alignment-oriented pre-training. The resulting tokens provide a compact and cacheable user representation that can be transferred to downstream rankers, decoupling lifelong-history representation learning from online candidate scoring. Experiments on the public KuaiRand benchmark and the industrial Tencent AdLive dataset, together with token-level analyses and a seven-day online A/B test, show that ChronicleRec consistently outperforms recent-window and single-pass compression baselines, recovers most of the performance of full attention with a compact cacheable representation, and delivers significant business gains in production.

\bibliographystyle{ACM-Reference-Format}
\bibliography{sample-base}

\appendix









\end{document}